\documentclass[pra,
                preprint,
                amsmath,
                amssymb,
                showpacs,
                superscriptaddress]{revtex4}

\usepackage{graphicx} 
\usepackage{dcolumn}  
\usepackage{bm}       
\usepackage{amsfonts}
\usepackage{dsfont}

\begin{document}

\title{Effect of screening on spectroscopic properties of Li-like ions in plasma environment}

\author{Pradip Kumar Mondal}
\affiliation{%
Department of Physics and Meteorology Indian Institute of
Technology Kharagpur,
            Kharagpur-721302, India }

\author{Narendra Nath Dutta}
\affiliation{%
Department of Physics and Meteorology Indian Institute of Technology Kharagpur,
            Kharagpur-721302, India }

\author{Gopal Dixit}
\email[]{gopal.dixit@cfel.de}
\affiliation{%
Center for Free-Electron Laser Science, DESY,
            Notkestrasse 85, 22607 Hamburg, Germany }

\author{Sonjoy Majumder}
\email[]{sonjoy@gmail.com}
\affiliation{%
Department of Physics and Meteorology Indian Institute of Technology Kharagpur,
            Kharagpur-721302, India }

\date{\today}

\begin{abstract}
\noindent This work presents accurate {\it ab initio}
investigations of various spectroscopic properties of a few
Li-like ions in presence of a plasma environment within the Debye
screening potential. The coupled-cluster theory in the
relativistic framework has been employed to compute ionization
potentials, excitation energies, electric dipole oscillator
strengths, and electric quadrupole transition probabilities of
Li-like C$^{3+}$, N$^{4+}$, and O$^{5+}$ ions. The unretarded
Breit interaction has been implemented to increase the accuracy of
the calculations. The effects of ion density and temperature on
the ionization potentials, excitation energies, electric dipole
oscillator strengths, and electric quadrupole transition
probabilities have been investigated in the plasma environment. It
is found that the plasma screening leads to a sharp decrease in
the ionization potential as the screening strength increases. With
increasing strength, the oscillator strengths associated with
2$s~^{2}S_{1/2}$$\rightarrow$2$p~^{2}P_{1/2, 3/2}$ transitions
increase, whereas the transition probabilities associated with
3$d$~$^2D_{3/2, 5/2}$$\rightarrow$2$s$~$^2S_{1/2}$ transitions
decrease.
\end{abstract}

\maketitle

\section{Introduction}
With the advent of novel x-ray sources based on free-electron
lasers (FELs)~\cite{emma2, ishikawa2012}, laser
plasmas~\cite{rousse2001}, and high-harmonic
generations~\cite{mckinnie2010, popmintchev2012}, it is possible
to achieve extreme conditions in matter such as high energy
density and high temperature using ultraintense, ultrashort, and
tunable pulses, and hence it is possible to create matter in
plasma form~\cite{vinko2012, ciricosta2012}. There have been many
experimental~\cite{murnane1991ultrafast, riley1992plasma,
mourou1992development, rogers1994astrophysical, nazir1996x,
workman1997application, nantel1998pressure, woolsey1998competing,
saemann1999isochoric, vinko2012, ciricosta2012} and theoretical~
\cite{rouse1971screening, gupta1982density, seidel1995energy,
ray1998magnetic, jung1999orientation, ray2000influence,
pang2002analytic, saha2002energy, okutsu2005electronic,
saha2005spherically, saha2006isoelectronic, li2008influence,
bhattacharyya2008effect, saha20092pnp, paul2009hydrogen,
qi2009generalized, sil2009spectra, gao2011plasma, xie2012energy}
endeavors to explain and understand the effect of plasma
environment on the spectroscopic properties of atoms and ions. In
the situation, when atoms or ions embedded in plasma, the
interaction between the nucleus and the bound electrons is
screened by the surrounding ions and fast electrons. The modified
interaction gives rise to phenomena such as pressure ionization and
continuum lowering and affects the spectroscopic properties
of atoms and ions~\cite{griem1974spectral, jaskolski1996confined}.
Recent advanced experiment, carried out using FEL and electron
beam ion trap (EBIT), provides an unexpected low oscillator
strength of electric dipole ($E1$) transition of Fe$^{16+}$ and
raises the concern about the quality of the atomic wave functions
used to model such spectral properties~\cite{bernitt2012}.
Therefore, treating the effect of plasma environment in atoms and
ions along with an accurate treatment of electron-electron
correlation and relativistic effects are nontrivial. The ratio of
Coulomb energy to thermal energy determine the strength of
coupling ($\Gamma$) in plasma. The low density and high
temperature situation corresponds to weakly coupled plasma
($\Gamma < 1$), where the screening of the nuclear Coulomb
interaction by free electrons in the plasma is guided by the Debye
model~\cite{ichimaru1982strongly, murillo1998dense}.

Lithium and lithium-like ions in plasma are a few of the most
abundant ionic species for specific temperature and density
attainable in the laboratory~\cite{kawachi1995}. Various
spectroscopic properties of Li-like ions have significant
importance in astrophysics due to evidence of high abundances of
these ions in different astronomical systems like active galactic
nuclei, x-ray binaries, quasars, and hot plasmas~\cite{nahar2000,
gorczyca2006}. For such small-sized atoms, allowed and forbidden
transitions with sufficient intensity are used as diagnostic tools
of tokamak plasmas~\cite{suckewer1981, godefroid1984, das1998}. It
is well known that the ionized form of carbon, nitrogen, and oxygen
and their various transition lines are important for the chromosphere
region of the solar atmosphere~\cite{fontenla2007}, and to understand
the dynamics and nature of the stellar and interstellar
medium~\cite{barstow2010, ferrero2011, prochaska2011, fox2011,
de2012}. There is long literature on the applications of the
isolated resonance lines having wavelength 1548.19 \AA~and 1550.77
\AA~for C$^{3+}$, 1238.82 \AA~and 1242.80 \AA~for N$^{4+}$ and
1031.91 \AA~and 1037.61 \AA~for O$^{5+}$~\cite{peach1988,
nist2012}. The astronomical observed lines are expected to be
affected by the plasma atmosphere at the origin and therefore
important for plasma diagnostic purpose. Forbidden transition
lines, i.e., electric quadrupole ($E2$) and magnetic dipole ($M1$)
transition lines provide very crucial parameters for estimations
of density and internal temperature measurements at low density
hot plasmas~\cite{burgess1998, allende2002, chen2010}. Also, the
transition rates of forbidden transitions provide accurate
dielectric recombination rates for these ions~\cite{andersen1992,
qu1999, laming2003}.

Several theoretical methods have been used to model the effect of
plasma environment on the spectroscopic properties for
one-electron~\cite{saha2002energy, ray2000influence,
bhattacharyya2008effect} and many-electron
~\cite{saha2006isoelectronic, li2008influence, das2011, das2012,
chaudhuri2012} systems. Due to the screening effect, lowering of
the ionization potential is demonstrated by Stewart and
Pyatt~\cite{stewart1966lowering}. The Debye plasma screening on
lighter atoms or ions have been studied over last decade using
different many-body approaches~\cite{bielinska2004relativistic,
saha2006isoelectronic, li2008influence, bhattacharyya2008effect}
and showed enough avenues of improvement. Recent works of
correlation exhaustive Dirac-Coulomb based coupled -cluster
calculations on He-, Li-, Be- and Na-like ions \cite{das2011,
das2012, chaudhuri2012} are examples of this. The authors of
these works emphasized the importance of relativistic correction
on the plasma screening by using more accurate many-body theories
within a relativistic framework~\cite{das2011, das2012,
chaudhuri2012}.

In the present article, we analyze the influence of the plasma
screening on the Li-like C$^{3+}$, N$^{4+}$, and O$^{5+}$ ions
using the Debye model potential. The ionization potentials,
excitation energies, oscillator strengths of $E1$ transitions, and
transition rates of $E2$ transitions are estimated for these ions
in the isolated (free) condition as well as within the plasma
environment. 
Here, we have used the Fock-space
coupled-cluster (FSCC) method within the Dirac-Coulomb-Breit (DCB)
Hamiltonian to consider the relativistic effect on these spectral
properties. Recently, Dutta $\it {et~al.}$ have implemented the unretarded
Breit interaction in an all-order approach using the
coupled-cluster (CC) theory and demonstrate the effect of
electron-electron correlation and unretarded Breit interaction on
the boron isoelectronic sequence~\cite{dutta2012}. It is
well known that the dynamical electron correlation, relaxation
effect, and Breit interaction are important for moderately charged
ions, which are considered here in an accurate way. This paper is
structured as follows. Section II discusses brief theory and
formalism of the relativistic FSCC approach with the Debye
screening potential. Section III presents results and discussions
on several spectral properties of Li-like ions and the effect of
plasma screening on these properties. Conclusions and future
outlooks are presented in Sec. IV.

\section{Theory}
In order to consider the effect of plasma environment on the
spectroscopic properties, the Dirac-Coulomb (DC) Hamiltonian with
unretarded Breit interaction for an $N$-electron atomic system can
be written as
\begin{equation}
H=\sum_{i=1}^{N}\left(c\overrightarrow{\alpha_i}\cdot\overrightarrow{p_i}+\left(
\beta_i-1\right) c^2 +V^{D}_{\textrm{eff}}(r_i)+
\sum_{j<i}\left(\frac{1}{r_{ij}}-\frac{\overrightarrow{\alpha_i}\cdot\overrightarrow{\alpha_j}}{r_{ij}}\right)\right),
\end{equation}
with all the standard notations often used. Here,
$V^{D}_{\textrm{eff}}(r_i)$ is the effective potential of the
nucleus on the $i$-th electron due to the presence of plasma
environment. The Debye-H\"uckel potential is considered to examine
the effect of screening of nuclear Coulomb potential due to the
presence of ions and free electrons in
plasmas~\cite{akhiezer1975plasma, ichimaru1982strongly}. In case
of weakly interacting plasma medium, the effective potential
experienced by the $i$-th electron is given as
\begin{equation}
V^{D}_{\textrm{eff}}(r_i)= -\frac{Ze^{-\mu r_i}}{r_i},
\end{equation}
where $Z$ is the nuclear charge and $\mu$ is the Debye screening
parameter, which is related to the ion density $n_{\textrm{ion}}$
and plasma temperature $T$  through the following relation:
\begin{equation}
\mu = \left[ \frac{4\pi (1+Z)n_{\textrm{ion}}}{k_{B}T}
\right]^{\frac{1}{2}},
\end{equation}
where $k_{B}$ is the Boltzmann constant. Therefore, a given value
of $ \mu $ represents a range of plasma conditions with different
ion densities and temperatures. The inverse of the Debye screening
parameter is called the Debye screening length, i.e., $\lambda_{D} =
\mu^{-1}$. The pure Coulomb nuclear attraction corresponds to the
zero screening situation ($\mu = 0$).

The wave functions, ionization potentials (IPs) of the ground, and
the different excited states for the considered $N$-electron
atomic system are obtained using the FSCC method with single,
double, and partial triple excitations within the relativistic
framework. The basic formalism of the FSCC method was developed
several decades before~\cite{lindgren1985, lindgren1987,
haque1984, pal1987, pal1988}. The relativistic version of the FSCC
theory has been developed recently and successfully employed to
obtain the various properties in different single valence atomic
systems~\cite{eliav1994, isaev2004, sur2006, dixit2007PRA,
dixitcd, dixitzn, pal2007relativistic, mani2010atomic}. Here, we
provide key steps of this method. In the relativistic
coupled-cluster (RCC) theory, the correlated wave function of a
single valence atomic state with a valance electron in the ``$v$"th
orbital is written in the form,
\begin{equation}
|\Psi_v\rangle=e^{T}\{1+S_v\}|\Phi_v\rangle
\end{equation}
where, $|\Phi_v\rangle$ is the corresponding reference state
generated at the Dirac-Fock (DF) level of the $N-1$ electron
closed-shell system using Koopman's theorem~\cite{szabo1996}. $T$
represents all possible excitations from the core orbitals of the
closed-shell system, and $S_v$ represents all possible valence and
core-valence excitations of the single-valence system. The detail
inclusion of the unretarded Breit interaction in this formalism is
described by Dutta and Majumder~\cite{dutta2012}.

The transition matrix element for any operator $O$ in the
framework of the RCC wave function can be expressed as
\begin{eqnarray}
O_{fi}& = & \frac{\langle \Psi_f|O|\Psi_i\rangle} {\sqrt{{\langle
\Psi_f|\Psi_f\rangle}
{\langle \Psi_i|\Psi_i\rangle}}} \nonumber \\
&=& \frac{{\langle
\Phi_f|\{1+{S_f}^{\dag}\}{e^T}^{\dag}Oe^T\{1+S_i\}|\Phi_i\rangle}}
{\sqrt{{\langle
\Phi_f|\{1+{S_f}^{\dag}\}{e^T}^{\dag}e^T\{1+S_f\}|\Phi_f\rangle}
{\langle
\Phi_i|\{1+{S_i}^{\dag}\}{e^T}^{\dag}e^T\{1+S_i\}|\Phi_i\rangle}}}.
\end{eqnarray}
The single-electron reduced matrix elements corresponding to $E1$,
$E2$, and $M1$ transitions are discussed in Refs.~\cite{grant1974,
johnson1995}.

The transition probabilities (in s$^{-1}$) corresponding to
$E1$, $E2$, and $M1$ channels from state {\it k} to {\it i} is
given as
\begin{equation}
A^{E1}_{k \rightarrow i} =
\frac{2.0261\times10^{18}}{\lambda^{3}g_{k}}S^{E1},
\end{equation}
\begin{equation}
A^{E2}_{k \rightarrow i} = \frac{1.11995\times10^{18}}{\lambda^{5}
g_{k}}S^{E2}
\end{equation}
and
\begin{equation}
A^{M1}_{k \rightarrow i} = \frac{2.69735\times10^{13}}{\lambda^{3}
g_{k}}S^{M1},
\end{equation}
where $S = {|{\langle \Psi_k|O|\Psi_i\rangle}|}^2$ is the
transition strength of the operator $O$ (in a.u.), $\lambda$ (in
\AA ) is the corresponding transition wavelength, and $g_{k} =
2j_k+1$ is the degeneracy of the $k$ state. The oscillator
strength of $E1$ transition from state {\it i} to {\it k} is given
as
\begin{equation}
f^{E1}_{i \rightarrow k} = 1.4992 \times 10^{-16}
\frac{g_{k}}{g_{i}} \lambda^{2} A^{E1}_{k \rightarrow i}.
\end{equation}

\section{Results and Discussions}
The precise DF orbital wave functions, which are the building block of
accurate correlation calculations, are generated via the basis set
expansion technique in the potential of two core electrons at
$1s_{1/2}$ orbital. The radial part of these basis wave functions
are considered to be Gaussian type~\cite{chaudhuri2000} having two
optimized exponential parameters $\alpha_0$ and $\beta$. The
nuclei are considered as finite size with a Fermi-type charge
distribution~\cite{parpia1992}. For all the ions considered here,
the number of basis wave functions at the DF levels for $s$-, $p$-,
$d$-, and $f$-type symmetries are 30, 25, 20, and 20, respectively.
In order to choose the optimized parameters for each ion, the
energies and wave functions of the DF orbitals are compared with
the same as obtained from the GRASP92 code, where the DF equations
are solved numerically~\cite{parpia2006}. These exponential
parameters are chosen as 0.005825 and 2.73 for C$^{3+}$, 0.003265
and 2.73 for N$^{4+}$, and 0.00525 and 2.73 for O$^{5+}$. The
number of DF orbitals for different symmetries used in the RCC
calculations are based on the convergent criteria of core
correlation energies with increasing number of orbitals. There are
12, 11, 10, and 10 active orbitals, which include all core
orbitals, considered in the calculations for the $s$, $p$, $d$,
and $f$ type symmetries, respectively. As an improvement of the
atomic Hamiltonian beyond the Dirac-Coulomb limit, the Breit
interaction in it's unretarded approximation has been included for
more accurate relativistic descriptions of the wave
functions~\cite{dutta2012}. The $T$ amplitudes are solved first
for the closed-shell systems and later the $S$ amplitudes
corresponding to different single-valence states are solved for
the open-shell systems using the RCC equations.

The quality of the wave functions for different eigen states is
ensured by comparing the reduced matrix elements of the $E1$
transitions in length and velocity gauges~\cite{johnson1995}. The
calculated average deviations of these matrix elements between
these two gauges at the CC levels have been found 1.11$\%$ for
C$^{3+}$, 1.17$ \% $ for N$^{4+}$, and 1.58$ \% $ for O$^{5+}$,
which indicate very good quality of the relativistic wave
functions. The calculated values of the ground state IPs for
C$^{3+}$, N$^{4+}$, and O$^{5+}$ are 520145, 789494 and 1113968
cm$^{-1}$, respectively; which are in excellent agreement with the
NIST results: 520178, 789537 and 1114004 cm$^{-1}$, respectively.
The average deviations of our calculated excitation energies (EEs)
of different excited states with respect to the NIST results are
estimated around 0.05$\%$, 0.06$\%$ and 0.06$\%$ for C$^{3+}$,
N$^{4+}$, and O$^{5+}$, respectively~\cite{nist2012}.

The $E1$ oscillator strengths are presented in
Table~\ref{tab:results1} along with the available NIST
results~\cite{nist2012} for comparison. The average deviations
between our calculated results and the NIST values are about 1.6\%
for C$^{3+}$, 1.2\% for N$^{4+}$, and 1.2\% for O$^{5+}$. Our
results for oscillator strength agree excellently with the recent
FSCC results presented by Das {\it et~al.} for C$^{3+}$
ion~\cite{das2012}. The recently calculated results of Elabidi et
al. using SUPERSTRUCTURE code~\cite{elabidi2011} have an average
deviation of about 3\% with respect to the NIST results; though,
in some cases deviations are around 25\%~\cite{elabidi2011}.
Therefore, correlation exhaustive and relativistic calculations
were wanted, and our calculations are motivated towards that. The
oscillator strengths of 3$d$~$^{2}D_{3/2,5/2}$$\rightarrow$
4$f$~${^2}F_{5/2,7/2}$ transitions are estimated here as well.

\begin{table}[h!]
\caption{Oscillator strengths of $E1$ transitions and their
comparisons with the NIST results for C$^{3+}$, N$^{4+}$, and
O$^{5+}$.} \label{tab:results1}
\begin{tabular}{lcrrrrrrr}
\hline\hline
\multicolumn{3}{c}{Terms} & \multicolumn{2}{c}{C$^{3+}$}& \multicolumn{2}{c}{N$^{4+}$} & \multicolumn{2}{c}{O$^{5+}$}\\
Lower &  & Upper & RCC & NIST & RCC & NIST & RCC & NIST\\
\hline
2$s$ $^2S_{1/2}$&$\rightarrow$&2$p$ $^2P_{1/2}$  & 0.095 &0.095 &0.078 &0.078 &0.066 & 0.066 \\
              &$\rightarrow$&2$p$ $^2P_{3/2}$  & 0.190 &0.190 &0.156 &0.156 &0.133 &0.133  \\
              &$\rightarrow$&3$p$ $^2P_{1/2}$  &0.067  &0.068 &0.078 &0.079 &0.087 &0.089  \\
              &$\rightarrow$&3$p$ $^2P_{3/2}$  & 0.133 &0.136 &0.156 &0.159 &0.174 &0.177  \\
              &$\rightarrow$&4$p$ $^2P_{1/2}$  &0.019  &0.020 &0.021 &0.023 &0.023 &0.025 \\
              &$\rightarrow$&4$p$ $^2P_{3/2}$  & 0.037 &0.041 &0.042 &0.046 &0.047 &0.049  \\
2$p$ $^2P_{1/2}$&$\rightarrow$&3$d$ $^2D_{3/2}$  &0.645  &0.646 &0.651 &0.652 &0.654 &0.657    \\
              &$\rightarrow$&4$d$ $^2D_{3/2}$  &0.122  &0.123 &0.122 &0.122 &0.120 &0.123    \\
2$p$ $^2P_{3/2}$&$\rightarrow$&3$d$ $^2D_{3/2}$  &0.064  & 0.065&0.065 &0.065 &0.065 & 0.066   \\
              &$\rightarrow$&3$d$ $^2D_{5/2}$  &0.581  &0.581 &0.586 &0.588 &0.590 &0.591    \\
              &$\rightarrow$&4$d$ $^2D_{3/2}$  &0.012  &0.012 &0.065 &0.065 &0.012 & 0.012   \\
3$s$ $^2S_{1/2}$&$\rightarrow$&3$p$ $^2P_{1/2}$  &0.161  &0.160 &0.131 &0.131 &0.112 &0.111   \\
              &$\rightarrow$&3$p$ $^2P_{3/2}$  &0.323  &0.320 &0.267 &0.263 &0.226 &0.224   \\
              &$\rightarrow$&4$p$ $^2P_{1/2}$  &0.063  &0.068 &0.077 &0.082 &0.092 &0.092   \\
              &$\rightarrow$&4$p$ $^2P_{3/2}$  &0.126  &0.136 &0.153 &0.164 &0.182 & 0.185  \\
3$p$ $^2P_{1/2}$&$\rightarrow$&3$d$ $^2D_{3/2}$  &0.062  &0.063 &0.054 & 0.055&0.049 &0.049    \\
              &$\rightarrow$&4$d$ $^2D_{3/2}$  &0.528  &0.541 &0.540 &0.550 &0.542 &0.557    \\
3$p$ $^2P_{3/2}$&$\rightarrow$&3$d$ $^2D_{3/2}$  &0.006  &0.006 &0.005 &0.005 &0.005 & 0.005   \\
              &$\rightarrow$&3$d$ $^2D_{5/2}$  & 0.055 &0.056 &0.048 &0.049 &0.043 & 0.044   \\
              &$\rightarrow$&4$d$ $^2D_{3/2}$  & 0.053 &0.054 &0.054 &0.055 &0.054 &  0.056  \\
              &$\rightarrow$&4$d$ $^2D_{5/2}$  &0.476  &0.486 &0.486 &0.495 &0.489 & 0.501   \\
3$d$ $^2D_{3/2}$&$\rightarrow$&4$f$ $^2F_{5/2}$  &1.020  & &1.020 &1.020 &1.020 & 1.010  \\
3$d$ $^2D_{5/2}$&$\rightarrow$&4$f$ $^2F_{5/2}$  &0.049  & &0.049 &0.048 &0.049 &0.048   \\
              &$\rightarrow$&4$f$ $^2F_{7/2}$  &0.974  & &0.975 &0.967 &0.974 &0.966   \\
\hline\hline
\end{tabular}\\
\end{table}

Table~\ref{tab:results2} presents the  emission probabilities of
$E2$ transitions having values of the order of 10$^4$ s$^{-1}$
or more using precisely calculated wavelengths for C$^{3+}$,
N$^{4+}$, and O$^{5+}$. Sur and Chaudhuri have reported a few $E2$
transition probabilities for O$^{5+}$ using the RCC theory based
on the DC Hamiltonian~\cite{sur2007}. Their results differ by
about 0.8\% from our calculated values obtained by the same theory
but based on the DCB Hamiltonian. All the transitions presented in
the Table~\ref{tab:results2} fall in the ultraviolet region of
electromagnetic spectrum. The wavelengths of these forbidden lines
relative to those of allowed lines from the same ion make them
very good candidates for line profile measurements and help to
understand the excitation processes, like electron- and
proton-impact excitations~\cite{feldman1981, dworetsky1998}. It is
evident from the table that there are few strong $E2$ transitions
with transition probabilities of the order of $10^{6}$ s$^{-1}$.
These are the transitions between the ground states and
3$d$~${^2}D_{3/2, 5/2}$ states for N$^{4+}$, and O$^{5+}$, and the
transitions 3$p$~$^{2}P_{1/2}$ $\rightarrow$ 2$p$~$^{2}P_{3/2}$ and
4$f$~$^{2}F_{5/2}$ $\rightarrow$ 2$p$~$^{2}P_{1/2}$ for O$^{5+}$.
Therefore, these forbidden transition lines having relatively
higher probabilities can play in the determinations of density and
internal temperature inside hot plasmas. Because the $M1$
transition probabilities are found to be quite low (of the order
of $ 10^{-1} $ s$^{-1}$ or less), the lines associated with
these transitions are hardly possible to detect and hence, are
excluded from the consideration here.

\begin{table}[h!]
\caption{Transition probabilities (in 10$^{4}$ s$^{-1}$) of $E2$
transitions along with corresponding transition wavelengths (in
\AA) for C$^{3+}$, N$^{4+}$, and O$^{5+}$.} \label{tab:results2}
\begin{tabular}{lcrrrrrrrrrr}
\hline\hline
\multicolumn{3}{c}{Terms} & \multicolumn{3}{c}{C$^{3+}$}& \multicolumn{3}{c}{N$^{4+}$} & \multicolumn{3}{c}{O$^{5+}$}\\
\hline
      &   &       & \multicolumn{2}{c}{$ \lambda $} &$A_{if}$ & \multicolumn{2}{c}{$ \lambda $} & $ A_{if} $ & \multicolumn{2}{c}{$ \lambda $} &$ A_{if} $\\
Upper &  & Lower  & RCC           &    NIST         &         & RCC        & NIST               &            & RCC          & NIST             &\\
\hline
3$d$ $^2D_{3/2}$&$\rightarrow$&2$s$ $^2S_{1/2}$  &307.71 &307.81 &44.03 &206.36 &206.43 &150.17 &148.16 &148.23 &413.17 \\
3$d$ $^2D_{5/2}$&$\rightarrow$&2$s$ $^2S_{1/2}$  &307.70 & 307.79&44.04 &206.35 &206.43 &150.23 &148.15 &148.21 &413.41\\
4$d$ $^2D_{3/2}$&$\rightarrow$&2$s$ $^2S_{1/2}$  &243.39 &243.71 &10.33 &161.62 &161.83 &30.06 &115.21 &115.35 &78.05\\
4$d$ $^2D_{5/2}$&$\rightarrow$&2$s$ $^2S_{1/2}$  &243.39 &243.71 &12.30 &161.61 &161.83 &30.09 &115.21 &115.35 &78.14\\
3$p$ $^2P_{3/2}$&$\rightarrow$&2$p$ $^2P_{1/2}$  &391.08 &391.23 &5.31 &251.52 &251.64 &20.01 &175.42 &175.47 &59.53\\
4$p$ $^2P_{3/2}$&$\rightarrow$&2$p$ $^2P_{1/2}$  &290.56 &290.83 &1.91 &186.81 &187.01 &7.19 &130.32 &130.37 &22.87\\
4$f$ $^2F_{5/2}$&$\rightarrow$&2$p$ $^2P_{1/2}$  &289.03 &289.05 &19.99 &185.99 &186.01 &76.16 &129.72 &129.75 &228.22\\
3$p$ $^2P_{1/2}$&$\rightarrow$&2$p$ $^2P_{3/2}$  &391.32 &391.45 &10.61 &251.76 &251.86 &40.01 &175.65 &175.68 &118.97\\
3$p$ $^2P_{3/2}$&$\rightarrow$&2$p$ $^2P_{3/2}$  &391.26 &391.40 &5.30 &251.71 & 251.81&19.99 & 175.61&175.63 &59.46\\
4$p$ $^2P_{1/2}$&$\rightarrow$&2$p$ $^2P_{3/2}$  &290.68 &290.93 &3.82 &186.92 &187.11 &14.35 &130.43 &130.47 &45.62\\
4$p$ $^2P_{3/2}$&$\rightarrow$&2$p$ $^2P_{3/2}$  &290.66 &290.92 &1.91 &186.91 &187.11 &7.18 &130.42 &130.46 &22.83\\
4$f$ $^2F_{5/2}$&$\rightarrow$&2$p$ $^2P_{3/2}$  &289.13 &289.14 &5.71 &186.09 &186.11 &21.76 &129.82 &129.84 &65.20\\
4$d$ $^2D_{3/2}$&$\rightarrow$&3$s$ $^2S_{1/2}$  &926.31 &930.34 &2.36 &615.27 &618.08 &8.57 &438.96 &440.83 &24.74\\
4$d$ $^2D_{5/2}$&$\rightarrow$&3$s$ $^2S_{1/2}$  &926.27 &930.30 &2.36 &615.23 &618.05 &8.57 &438.92 &440.79 &24.74\\
4$p$ $^2P_{3/2}$&$\rightarrow$&3$p$ $^2P_{1/2}$  &1130.07 &1132.83 &0.54 &725.71 &727.75 &2.02 &506.47 &506.88 &6.18\\
4$f$ $^2F_{5/2}$&$\rightarrow$&3$p$ $^2P_{1/2}$  &1107.17 &1106.38 &2.22 &713.41 &712.79 &8.29 &497.54 &497.61 &24.40\\
4$p$ $^2P_{1/2}$&$\rightarrow$&3$p$ $^2P_{3/2}$  &1130.75 &1133.41 &1.07 &726.35 &728.33 &4.04 &507.10 &507.44 &12.36\\
4$p$ $^2P_{3/2}$&$\rightarrow$&3$p$ $^2P_{3/2}$  &1130.54 &1133.24 &0.54 &726.16 &728.16 &2.02 &506.91 &507.28 &6.18\\
4$f$ $^2F_{5/2}$&$\rightarrow$&3$p$ $^2P_{3/2}$  &1107.62 &1106.77 &0.63 &713.84 &713.86 &2.36 &497.96 &497.98 &6.95\\
4$f$ $^2F_{7/2}$&$\rightarrow$&3$p$ $^2P_{3/2}$  &1107.61 &1106.77 &2.85 &713.82 &713.15 &10.64 &497.94 &497.96 &31.29\\
4$d$ $^2D_{3/2}$&$\rightarrow$&3$d$ $^2D_{3/2}$  &1164.01 &1170.18 &0.33 &745.01 &748.99 &1.27 &517.81 &520.14 &3.80\\
4$d$ $^2D_{5/2}$&$\rightarrow$&3$d$ $^2D_{3/2}$  &1163.95 &1170.13 &0.095 &744.95 &748.94 &0.36 &517.75 &520.08 &1.08\\
4$d$ $^2D_{3/2}$&$\rightarrow$&3$d$ $^2D_{5/2}$  &1164.13 &1170.33 &0.14 &745.13 &749.11 &0.55 &517.93 &520.28 &1.63\\
4$d$ $^2D_{5/2}$&$\rightarrow$&3$d$ $^2D_{5/2}$  &1164.08 & 1170.27&0.38 &745.08 &749.06 &1.41 &517.88 &520.22 &4.34\\
\hline\hline
\end{tabular}\\
\end{table}

\begin{figure}
\includegraphics[width=12cm]{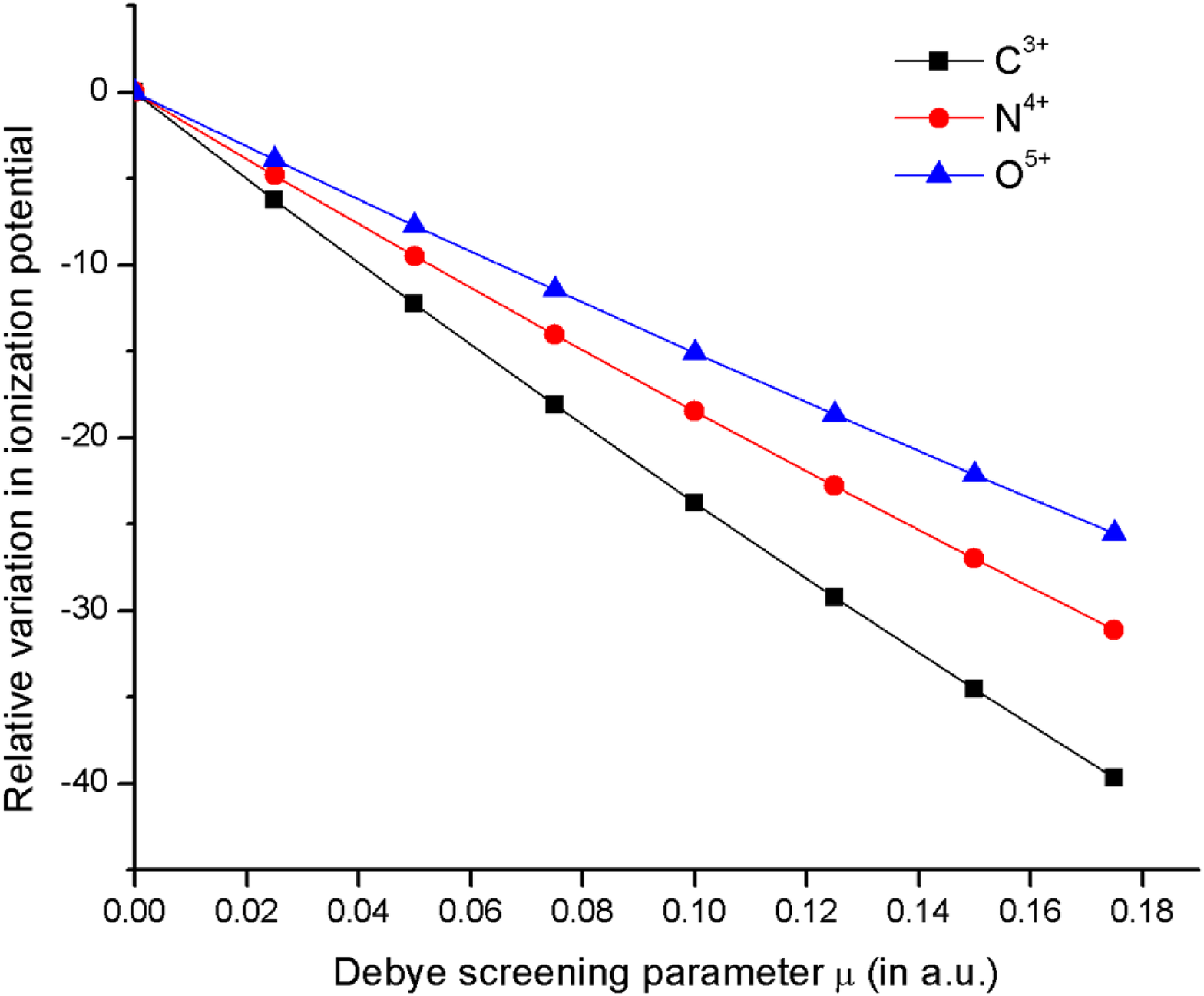}
\caption{(Color online) Relative variations of ionization
potentials with Debye screening parameter $\mu$ for C$^{3+}$,
N$^{4+}$, and O$^{5+}$.} \label{fig1}
\end{figure}

After validating the quality and accuracy of our method and
calculations for unscreened ions, i.e., with $\mu = 0$, several
spectroscopic properties for Li-like ions are calculated for
different values of Debye screening parameter $\mu$. In most of
the scenario, the ions are presented in low density plasmas where
all the spectroscopic properties are affected by the plasma
atmosphere which are quantified by the Debye screening parameter
$\mu$. We have chosen the values of $\mu$ ranging from 0 to 0.175
a.u. in interval of 0.025 a.u.. The ions become unstable after
$\mu = 0.175$ a.u.. It is mentioned already in the ``Theory"
ection that $\mu$ is a function of ion density $n_{\textrm{ion}}$
and plasma temperature $T$. Therefore, the above ranges of $\mu$
mimic weakly coupled plasma; for example, $T = 10^{6}~\textrm{K}$
and $n_{\textrm{ion}} \sim 10^{22}~{\textrm{cm}^{-3}}$ correspond
to $\mu = 0.15$ a.u.~\cite{das2012,sahoo2006}. This type of
condition can be achieved in the laboratory plasmas for high
temperature~\cite{sahoo2006}.  The relative variations in the IPs
as a function of $\mu$ for C$^{3+}$, N$^{4+}$, and O$^{5+}$ ions
are presented in Fig.~\ref{fig1}. For convenience, we define
relative variation in any spectroscopic property of interest here,
say $\textbf{O}$, as
\begin{equation}
\textbf{Relative variation in O} = \frac{[\textbf{O}(\mu \neq 0 )
- \textbf{O}(\mu = 0)] \times 100}{\textbf{O}(\mu = 0)}.
\end{equation}
It is evident from the Fig.~\ref{fig1} that as the ion density
increases or temperature decreases, i.e., the screening strength
$\mu$ increases, IPs decrease linearly and the systems become less
and less stable. This is because, the screening of the nuclear
charge increases with the increase of $\mu$ and hence, the
attractive nuclear Coulomb potential at the valence electron
decreases. This particular fact can be attributed as {\it
continuum lowering} for the system surrounded in plasma
environment. A similar trend was observed already for Li and
Li-like ions in the presence of Debye plasma
environment~\cite{sahoo2006, das2012}. In addition, this figure shows
that the fall in IP decreases with increasing ionic charge, i.e.,
the IP of C$^{3+}$ decreases more rapidly than the IPs of
N$^{4+}$, and O$^{5+}$ within a same interval of $\mu$. With
increasing nuclear charge $Z$ of an isoelectronic sequence, the
valence electron comes closer to the nucleus and hence can
defend more effectively the screening of the plasma environment.

\begin{figure}
\includegraphics[width=12cm]{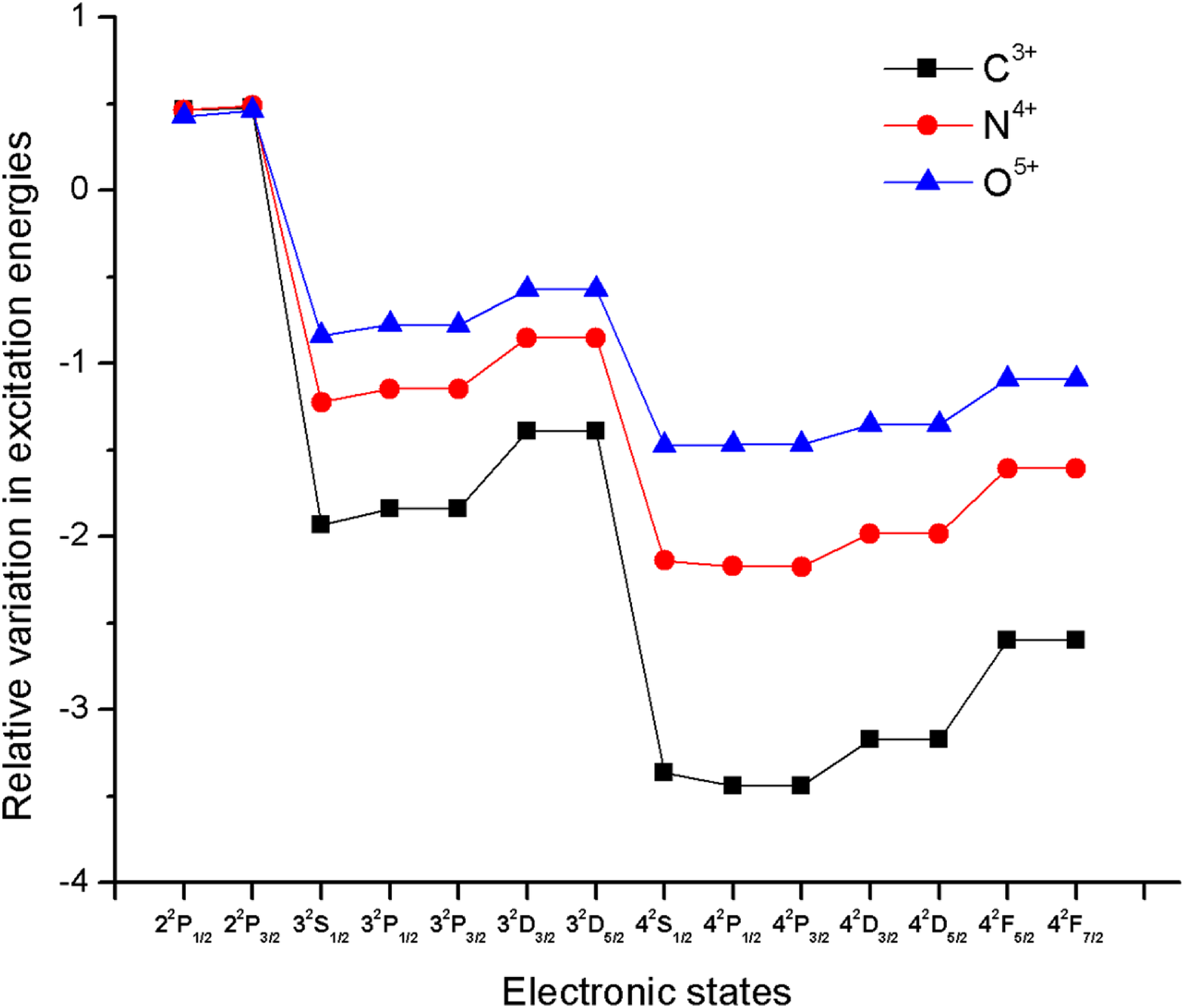}
\caption{(Color online) Relative variations of excitation energies
 of different excited states with Debye screening parameter $\mu = 0.075$ a.u. for
C$^{3+}$, N$^{4+}$, and O$^{5+}$.} \label{fig2}
\end{figure}

Figure~\ref{fig2} presents the relative variations in the
excitation energies (EEs) of different low-lying excited states
for $\mu = 0.075$ a.u. for the Li-like ions. For this value of
$\mu$, the effect of screening on the EEs is found optimum.
However, we have observed similar trends in the EEs for any other
values of $\mu$ within the given range of weakly coupled plasma.
It is evident from this figure that due to the screening, the
ground-state transitions from the states 2$p$~${^2}P_{1/2,3/2}$ are
blue-shifted, and from the others are red-shifted. This figure
further reflects that the shift in the EEs decreases as the
nuclear charge increases. These shifts in the EEs can be
attributed by the quantum confinement and electron screening in
the presence of the plasma environment~\cite{li2008influence}. It is well
known that the quantum defect decreases as the orbital angular
momentum quantum number increases, and for angular momentum
quantum number equal and larger than two, the quantum defect is
almost zero. Therefore, electronic states with higher angular
momentum quantum number among the same principle quantum number
experience relatively less effect of $\mu$. For example, EEs for
3$d$~$^{2}D_{3/2,5/2}$ are less perturbed than the states
3$s$~$^{2}S_{1/2}$ and 3$p$~$^{2}P_{1/2,3/2}$ due to screening. In addition,
the effect of quantum confinement is the same for the fine structure
states. The figure further confirms that the quantum confinement
is more pronounced for higher excited states than relatively
low-lying states~\cite{li2008influence}.

\begin{figure}
\includegraphics[width=17cm]{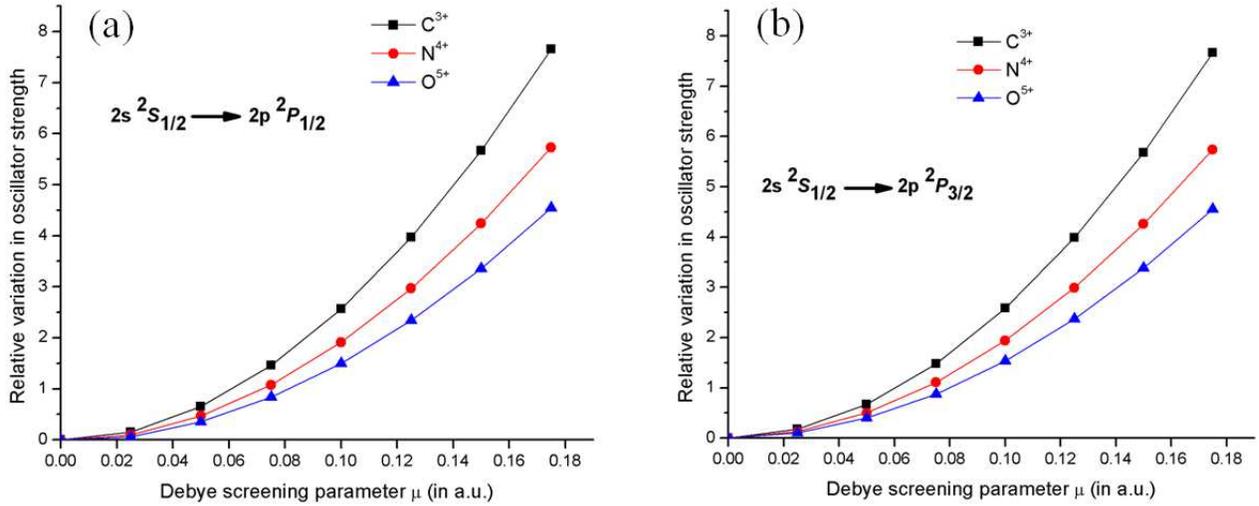}
\caption{(Color online) Relative variations of oscillator
strengths of $E1$ transitions with Debye screening parameter $\mu$
for C$^{3+}$, N$^{4+}$, and O$^{5+}$. Panels (a) and (b)
correspond to ${2s}~^{2}S_{1/2} \rightarrow
{2p}~^{2}P_{1/2}$ and ${2s}~^{2}S_{1/2} \rightarrow
{2p}~^{2}P_{3/2}$ transitions, respectively.} \label{fig3}
\end{figure}

The relative variations of oscillator strengths with respect to
$\mu$ values for the most strong 2$s$~$^{2}S_{1/2}$ $\rightarrow$
2$p$~$^{2}P_{1/2,3/2}$ transitions of C$^{3+}$, N$^{4+}$, and
O$^{5+}$ ions are shown in Fig.~\ref{fig2}. The relative
variations increase monotonically as $\mu$ increases and the
effect is the same for both the transitions. For a particular value of
$\mu$, the screening effect on the oscillator strengths decreases
as nuclear charge increases. However, the transition energies
corresponding to these transitions change significantly, which
reflect in the total change in the oscillator strengths as
observed in Fig.~\ref{fig3}. This present trend of the $E1$
oscillator strengths in Debye plasma has been reported in the
recent past~\cite{xie2012energy, das2012}. Our present findings
show that the lifetimes of 2p~$^{2}P_{1/2,3/2}$ states decrease
with increasing plasma strength $\mu$. The lifetimes of these
states depend on the third power of wavelengths of the associated
$E1$ transitions to the ground state. This enhances the screening
effect on the lifetimes with respect to the oscillator strengths
which depend inversely on the first power of the corresponding
wavelengths.

\begin{figure}
\includegraphics[width=15cm]{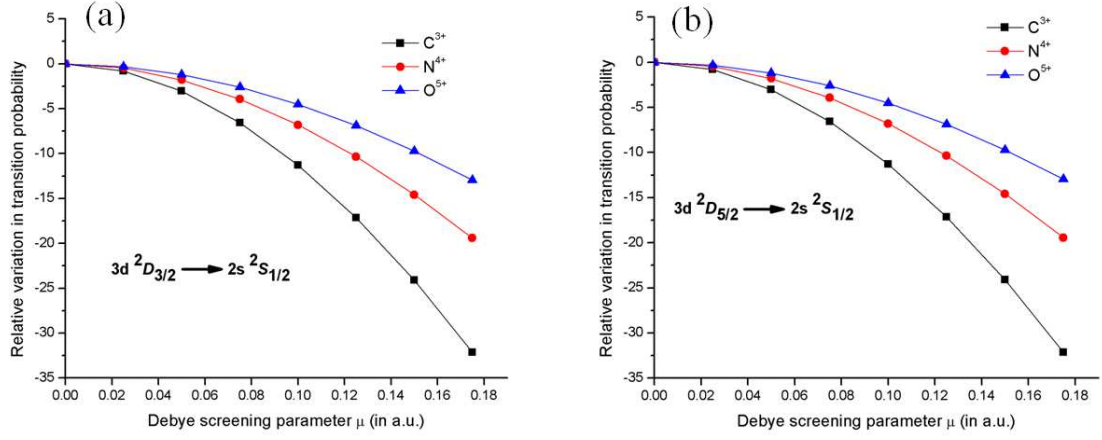}
\caption{(Color online) Relative variations of transition
probabilities of $E2$ transitions with Debye screening parameter
$\mu$ for C$^{3+}$, N$^{4+}$, and O$^{5+}$. Panels (a) and (b)
correspond to ${3d}~^{2}D_{3/2} \rightarrow
{2s}~^{2}S_{1/2}$, and (b) ${3d}~^{2}D_{5/2}
\rightarrow {2s}~^{2}S_{1/2}$ transitions, respectively.}
\label{fig4}
\end{figure}

The transition probabilities ($A^{E2}$) of the $E2$ transition
3$d$~$^{2}D_{3/2}$$\rightarrow$ 2$s$~$^2S_{1/2}$ along with the
corresponding transition wavelengths for different values of $\mu$
are presented in Table~\ref{tab:results3}. It is evident from the
table that the $E2$ transition wavelengths are red shifted with
increasing value of the screening strength. A similar trend is
observed for the 3$d$~$^{2}D_{5/2}$$\rightarrow$ 2$s$~$^2S_{1/2}$
transition, which is not presented here. The relative variations
in the transition probabilities for both these $E2$ transitions as
a function of $\mu$ are depicted in Fig.~\ref{fig4}. Both these
transitions are chosen due to their comparatively higher
probabilities with respect to other transitions as evident from
Table II. The systematic decrease of the transition
probabilities with increasing $\mu$ is observed from the figure.
One can also visualize that the relative decrease in $E2$
transition probabilities for a particular value of screening
parameter is more for C$^{3+}$ compared to O$^{5+}$. If one
compares the relative changes in $E1$ versus $E2$ transition
amplitudes as a function of $\mu$, one finds that the latter is
more influenced by $\mu$ than the former. This behavior can be
explained from their radial dependence. The amplitude of $E1$
transition has $r$ dependence, where that of $E2$ transition has
$r^2$ dependence. Due to this, $E2$ transition amplitude depends on
further field region compared to $E1$ transition amplitude from the
nucleus and hence, $E2$ transition amplitude is relatively more
affected by the screening. At this point, it is important to
emphasis that $E2$ transitions are the effective processes in the
low density hot plasmas. Therefore, estimations of the influence
of nuclear charge screening in the $E2$ transition probabilities
along with their wavelengths are of great importance in plasma
modeling.

\begin{table}[h!]
\caption{Effect of $\mu$ (in a.u.) on
$A^{E2}_{{3d}~^{2}D_{3/2} \rightarrow {2s}~^{2}S_{1/2}}$ (in 10$^{4}$ s$^{-1}$) and
corresponding transition wavelengths (in \AA) for C$^{3+}$, N$^{4+}$, and O$^{5+}$.}\label{tab:results3}
\begin{tabular}{lrrrrrr}
\hline\hline
$\mu$ & \multicolumn{2}{c}{C$^{3+}$}& \multicolumn{2}{c}{N$^{4+}$} & \multicolumn{2}{c}{O$^{5+}$}\\
      &   $ \lambda $ & $ A^{E2} $  & $ \lambda $ & $ A^{E2} $ & $ \lambda $ & $ A^{E2}$  \\
\hline
0     &   307.71      &    44.03    &  206.36     &  150.17    & 148.16      &413.17\\
0.025 &    308.22     &    43.68    &   206.57    & 149.47     & 148.26      &411.92\\
0.05  &    309.69     &    42.70    &   207.18    &  147.46    & 148.56      &408.29\\
0.075 &    312.10     &     41.14   &   208.17    &   144.26   & 149.04      &402.46\\
0.1   &   315.44      &    39.06    &   209.53    &   139.95   & 149.70      &394.56\\
0.125 &    319.75     &    36.48    &   211.27    &  134.62    & 150.54      &384.72\\
0.15  &    325.13     &    33.42    &   213.40    &  128.27    & 151.56      &373.04\\
0.175 &     331.71    &    29.87    &   215.94    &  121.00    & 152.76      &359.60\\
\hline\hline
\end{tabular}\\
\end{table}

\section{Conclusion}
We have investigated the influence of Debye screening of nuclear
charges due to the presence of free electrons and ions in plasma
medium on the ionization potentials, excitation energies, $E1$
oscillator strengths, and $E2$ transition probabilities of
C$^{3+}$, N$^{4+}$, and O$^{5+}$. Especially, the study on $E2$
transitions can be considered a useful tool to model low density
and high temperature plasmas. The transition wavelengths are
mainly affected by this screening, which characterizes the
screening effects on the other associated spectroscopic properties
like the oscillator strengths and transition probabilities. Due to
the high abundances of these Li-like ions in various astrophysical
systems, we hope our investigations will be useful to the
astrophysicist in the near future.

\begin{acknowledgments}
Pradip Kumar Mondal and Narendra Nath Dutta
recognize financial help from the Council of Scientific and
Industrial Research (CSIR), India.
\end{acknowledgments}

\end{document}